\documentclass[12pt]{iopart}
\usepackage{epsfig}

\newcommand{\be}{\begin{equation}}
\newcommand{\ee}{\end{equation}}
\newcommand{\bea}{\begin{eqnarray}}
\newcommand{\eea}{\end{eqnarray}}

\newcommand{\pp}{\,\, .}
\newcommand{\vv}{\,\, ,}

\def\fun#1#2{\lower3.6pt\vbox{\baselineskip0pt\lineskip.9pt
\ialign{$\mathsurround=0pt#1\hfill##\hfil$\crcr#2\crcr\sim\crcr}}}

\newcommand\gsim{\mathrel{\rlap{\lower4pt\hbox{\hskip1pt$\sim$}}
    \raise1pt\hbox{$>$}}}

\def\dslash{\not{\hbox{\kern-2pt $\partial$}}}
\def\Dslash{\not{\hbox{\kern-4pt $D$}}}
\def\Oslash{\not{\hbox{\kern-4pt $O$}}}
\def\Qslash{\not{\hbox{\kern-4pt $Q$}}}
\def\pslash{\not{\hbox{\kern-2.3pt $p$}}}
\def\kslash{\not{\hbox{\kern-2.3pt $k$}}}
\def\qslash{\not{\hbox{\kern-2.3pt $q$}}}

 \newtoks\slashfraction
 \slashfraction={.13}
 \def\slash#1{\setbox0\hbox{$ #1 $}
 \setbox0\hbox to \the\slashfraction\wd0{\hss \box0}/\box0 }

\def\ee{\end{equation}}
\def\be{\begin{equation}}

\begin{document}
\setlength{\unitlength}{1mm}

\title[Present bounds on the relativistic energy density]{Present bounds on the relativistic energy density in the Universe
from cosmological observables}

\author{Gianpiero Mangano\,$^1$, Alessandro Melchiorri \,$^2$, Olga Mena\,$^2$, \\ Gennaro Miele\,$^1$  and Anze
Slosar\,$^3$
\footnote[3]{mangano@na.infn.it\\
Alessandro.Melchiorri@roma1.infn.it
\\ Olga.MenaRequejo@roma1.infn.it\\ miele@na.infn.it\\
axs@astro.ox.ac.uk}}

\address{$^1$ \, Physics Department and Sezione INFN, University
of Naples ``Federico II'', \\ Via Cintia, 80126 Naples, Italy}

\address{$^2$\, Physics Department and Sezione INFN, University of Rome ``La
Sapienza'', \\ P.le Aldo Moro 2, 00185 Rome, Italy}

\address{$^3$ \, Astrophysics, Denys Wilkinson Building, University of
Oxford,\\ Keble Road, OX3RH1, Oxford, UK}


\begin{abstract}

We discuss the present bounds on the relativistic energy density
in the Universe parameterized in terms of the effective number of
neutrinos $N_\nu^{\rm \it eff}$ using the most recent cosmological
data on Cosmic Microwave Background (CMB) temperature anisotropies
and polarization, Large Scale galaxy clustering from the Sloan
Digital Sky Survey (SDSS) and 2dF, luminosity distances of type Ia
Supernovae, Lyman-$\alpha$ absorption clouds (Ly-$\alpha$), the
Baryonic Acoustic Oscillations (BAO) detected in the Luminous Red
Galaxies of the SDSS and finally, Big Bang Nucleosynthesis (BBN)
predictions for $^4$He and Deuterium abundances. We find
$N_\nu^{\rm \it eff}= 5.2^{+2.7}_{-2.2}$ from CMB and Large Scale
Structure data, while adding Ly-$\alpha$ and BAO we obtain
$N_\nu^{\rm \it eff}= 4.6^{+1.6}_{-1.5}$ at 95 \% c.l.. These
results show some tension with the standard value $N_\nu^{\rm \it
eff}=3.046$ as well as with the BBN range $N_\nu^{\rm \it eff}=
3.1^{+1.4}_{-1.2}$ at 95 \% c.l., though the discrepancy is
slightly below the 2-$\sigma$ level. In general, considering a
smaller set of data weakens the constraints on $N_\nu^{\rm \it
eff}$. We emphasize the impact of an improved upper limit (or
measurement) of the primordial value of $^3$He abundance in
clarifying the issue of whether the value of $N_\nu^{\rm \it eff}$
at early (BBN) and more recent epochs coincide.

\end{abstract}


\maketitle


\section{Introduction}

The energy density in relativistic species $\rho_{\rm R}$ is one of the
key parameters defining a specific cosmological model and is
usually parameterized in terms of the effective neutrino number
$N_\nu^{\rm \it eff}$ \be \rho_{\rm R} = \left[1+ \frac{7}{8} \left(
\frac{T_\nu}{T}\right)^{4} N_\nu^{\rm \it eff}\right] \rho_\gamma(T) \vv
\label{neff} \ee with $\rho_\gamma(T)$ the photon energy density
and $T_\nu$ the neutrino temperature. There is a very neat result
for its value after $e^+-e^-$ annihilation, namely
$N_\nu^{\rm \it eff}=3.046$ assuming standard electroweak interactions,
three active neutrinos and including the (small) effect of
neutrino flavor oscillations \cite{pinto}. On the other hand,
$N_\nu^{\rm \it eff}$ which in complete generality is a function of the
particular stage of the expanding Universe, i.e. of the photon
temperature, can be indirectly constrained by several cosmological
observables. The final yields of both $^4$He and D produced during
Big Bang Nucleosynthesis (BBN) are sensitive probes of the total
amount of radiation energy density which drives the expansion rate
via Friedmann equation, as well as of the neutrino phase space
distribution which enters the weak rates governing the eventual
neutron to proton number density ratio. At later stages, the
imprint of different values of $N_\nu^{\rm \it eff}$ can be extracted from
the temperature and polarization maps of photons during the
formation at the last scattering surface of the Cosmic Microwave
Background (CMB) or the power spectrum of Large Scale Structures
(LSS), mainly because its role in shifting the matter-radiation
equality and changing the early Integrated Sachs-Wolfe effect. For
recent reviews on these issues see e.g. \cite{Serpico:2004gx} and
\cite{Lesgourgues:2006nd}.

The huge amount of observations from CMB satellite, balloon-borne
and ground based experiments,
\cite{wmap3cosm}-\cite{2005astro.ph..7503M}, galaxy redshift
surveys, \cite{2004ApJ...606..702T,2005MNRAS.362..505C}, distant
type Ia Supernovae \cite{2006A&A...447...31A}, to only mention
those of larger impact, is pointing out a rather consistent {\it
standard} Cosmological Model (see however, \cite{Blanchard:2003du}
for a different viewpoint). On the other hand, all these data can
be fruitfully used to constrain {\it non standard} physics at the
fundamental level, such as classes of elementary particle models
predicting a different radiation content in the Universe.

Indeed, there are several theoretical models which have been
considered to account for different values of $N_\nu^{\rm \it eff}$ at
different stages as the physics involved at the BBN and at the CMB
epochs could be different~\cite{Hansen:2001hi,Bowen:2001in}, see
e.g. \cite{kaplinghat:2000jj} for the baryon energy density values
inferred from BBN and CMB data sets separately. Many of these
models involve additional sterile neutrino species which either
are generated in presence of a large lepton number and their
production is therefore suppressed avoiding constraints from CMB
and LSS, or are produced in the decay of an active neutrino
together with a majoron~\cite{White:1994as}. These models, in
general, assume a (light) massive sterile neutrino, motivated by
the LSND oscillation claim~\cite{Aguilar:2001ty}. However, current
cosmological data excludes at a high significant level the sterile
neutrino hypothesis as an explanation for the LSND signal
\cite{Dodelson:2005tp}. Nevertheless, it remains still viable to
enlarge the neutrino sector with massless sterile species by
imposing a discrete symmetry which prevents the extra sterile
neutrinos to acquire masses. For example, it has been recently
suggested in \cite{cgd} a {\it Leptocratic} model, which enlarges
the Standard Model by including a non-anomalous $U(1)_\nu$
symmetry and predicting the existence of light ($m_\nu \ll 1$ eV)
quasi-sterile neutrinos. Indeed, from a theoretical point of view,
all the models which contain massless or very light sterile
neutrinos pose a major challenge, namely to understand the sterile
neutrino lightness, despite being a Standard Model singlet. An
interesting possibility is the model discussed in
\cite{Chacko:2003dt}, with neutrinos acquiring a mass from a
broken lepton flavor symmetry via operators of the form ${\cal O}
= \ell \, N \, h \left({\phi / M}\right)^p$, where $\ell$ is the
left-handed lepton doublet, $h$ is the Standard Model Higgs field,
$M$ a mass scale larger that the electroweak scale, $p$ a positive
integer and finally, $N$ is a right handed neutrino. Such a model
contains a number of pseudo-Goldstone bosons ($\phi$) which could
modify the relativistic energy density at the eV era (CMB) while
keeping BBN essentially unchanged if these extra particles only
interact with Standard Model particles via operators of the form
${\cal O}$. This model have predictions for $N_\nu^{\rm \it eff}$ at the
CMB epoch in the range $3 \div 4$. For a recent combined study
constraining the Yukawa couplings of neutrino to low mass (pseudo)
scalars, see \cite{Hannestad:2005ex}.

Another possible scenario is to allow for a violation of the
spin-statistics theorem in the neutrino sector, so that the
neutrino distribution function interpolates between the fermionic
and the bosonic results, increasing the number of degrees of
freedom for CMB from $0$ up to $3/7$, due to the presence of
bosonic neutrinos \cite{Dolgov:2005mi}.

A different possibility is finally, to consider either a
quintessence field with a tracking potential behavior or an extra
interaction among the dark energy and radiation (or dark matter)
sectors or even a Brans-Dicke field with non trivial potential
\cite{DeFelice:2005bx} which could mimic the effect of adding
extra radiation between the BBN and CMB epochs. Indeed, we
presently know very little about the role of any possible dark
energy between BBN and CMB eras, despite the fact that the amount
of dark energy at both epochs is very well constrained
\cite{Bean:2001wt}.

These numerous theoretical studies, also motivated by the
improvement of the data accuracy which is expected from the next
generation of planned experiments like \verb"PLANCK", shows the
fact that there is quite an intense research activity which
intrigues several scholars and try to answer the following two
main questions
\begin{itemize}
\item[i)] what is the value of $N_\nu^{\rm \it eff}$ which is favored by
cosmological observations with respect to the standard result?
\item[ii)] is there any hint for a time dependent value of
$N_\nu^{\rm \it eff}$, which takes different values at, say, the BBN epoch
(photon temperature of order MeV) and CMB last scattering (1 eV)?
\end{itemize}

In this paper we address both these issues. Though several
analysis of the kind are already present in the literature
\cite{Hansen:2001hi}, \cite{Hannestad:2000hc}-\cite{hannestud}, we
provide updated results using a broad set of the most recent
available data from CMB, LSS, as well as using Baryon Acoustic
Oscillations (BAO) detection of \cite{2005ApJ...633..560E}, small
scale spectrum from Lyman-$\alpha$ forest
\cite{McDonald:2004eu,McDonald:2004xn} and finally, reconsidering
the constraints from BBN in view of a new Deuterium measurement
reported in \cite{O'Meara:2006mj} and using an improved BBN
numerical code described in \cite{Serpico:2004gx}. In particular,
our study has been stimulated by a a recent paper
\cite{Seljak:2006bg} where a mild disagreement (at the
2.4-$\sigma$ level) is found between the (large) value of
$N_\nu^{\rm \it eff}$ at late epochs (CMB and LSS) and both the standard
value and that singled out by BBN. Section II is devoted to a
summary of the data analysis method, while in Section III we
discuss our results. Finally, we draw our conclusions and outlooks
in Section IV.

\section{Data analysis method}

The method we adopt is based on the publicly available Markov
Chain Monte Carlo package \texttt{cosmomc} \cite{Lewis:2002ah}
with a convergence diagnostics done through the Gelman and Rubin
statistic. We sample the following seven-dimensional set of
cosmological parameters, adopting flat priors on them: the baryon
and Cold Dark Matter densities, $\omega_b=\Omega_bh^2$ and
$\omega_c=\Omega_ch^2$, the ratio of the sound horizon to the
angular diameter distance at decoupling, $\theta_s$, the scalar
spectral index $n_S$, the overall normalization of the spectrum
$A$ at $k=0.05$ Mpc$^{-1}$, the optical depth to reionization,
$\tau$ and the effective number of massless neutrinos
$N^{\rm \it eff}_\nu$.

We consider two sets of cosmological data. In the first set, we
include the three-year WMAP data \cite{wmap3cosm} (temperature and
polarization) with the routine for computing the likelihood
supplied by the WMAP team. Together with the WMAP data we also
consider the small-scale CMB measurements of CBI
\cite{2004ApJ...609..498R}, VSA \cite{2004MNRAS.353..732D}, ACBAR
\cite{2002AAS...20114004K} and BOOMERANG-2k2
\cite{2005astro.ph..7503M}. In addition to the CMB data, we
include the constraints on the real-space power spectrum of
galaxies from the SLOAN galaxy redshift survey (SDSS)
\cite{2004ApJ...606..702T} and 2dF \cite{2005MNRAS.362..505C}, and
the Supernovae Legacy Survey data from \cite{2006A&A...447...31A}.
Finally we include a prior on the Hubble parameter from the Hubble
Space Telescope Key project \cite{freedman}. In the second dataset
we add to the data quoted above the constraints from the Baryonic
Acoustic Oscillations (BAO) detected in the Luminous Red Galaxies
(LRG) sample of the SDSS \cite{2005ApJ...633..560E} and  we
include measurements of the small scale primordial spectrum from
Lyman-$\alpha$ forest clouds
\cite{McDonald:2004eu,McDonald:2004xn}.

The first dataset is based on well-established data while the
second include new results that, especially in the case of
Lyman-$\alpha$ data, should still be confirmed by further
experimental data. Therefore, even if more stringent the results
obtained from the second dataset should be considered as less
conservative. In the framework of the cosmological model and using
the data described above we produce constraints on $\omega_b$ and
$N^{\rm \it eff}_\nu$, after marginalization over the remaining 'nuisance'
parameters.

Constraints on these two parameters are also computed using
standard BBN theoretical prediction as provided by the new
numerical code described in \cite{Serpico:2004gx}, which includes
a detailed treatment of the three active neutrino decoupling
stage, as well as a full updating of all rates entering the
nuclear chain based on the most recent experimental results on
nuclear cross sections. We adopt for the D/H abundance ratio the
new average of \cite{O'Meara:2006mj} obtained including a new
measurement in a metal poor damped Lyman-$\alpha$ system along the
line of sight of QSO SDSS1558-0031 \be
\textrm{D/H}=(2.82_{-0.25}^{+0.27}) \cdot 10^{-5} \vv \label{deut}
\ee We use the uncertainty as quoted in \cite{O'Meara:2006mj},
computed by a jackknife analysis. Indeed, the six individual D/H
measurements used in their study show a scatter around the
weighted mean. This might be a real effect due to non homogeneous
D/H abundance, or rather the effect of underestimated errors in
measurements.

Finally, we also use the conservative range for the $^4$He mass
fraction $Y_p$ obtained in \cite{olive} \be Y_{p} = 0.249 \pm
0.009 \pp \label{4he} \ee The likelihood function is then
constructed as in \cite{Serpico:2004gx}, by also taking into
account the (small) D - $^4$He correlation term in the inverse
covariance matrix due to the effect of the propagated uncertainty
of the several nuclear rates on the theoretical nuclide
abundances.

\section{Results}

\begin{figure}

\begin{tabular}{cc}
  \includegraphics[width=8.1cm]{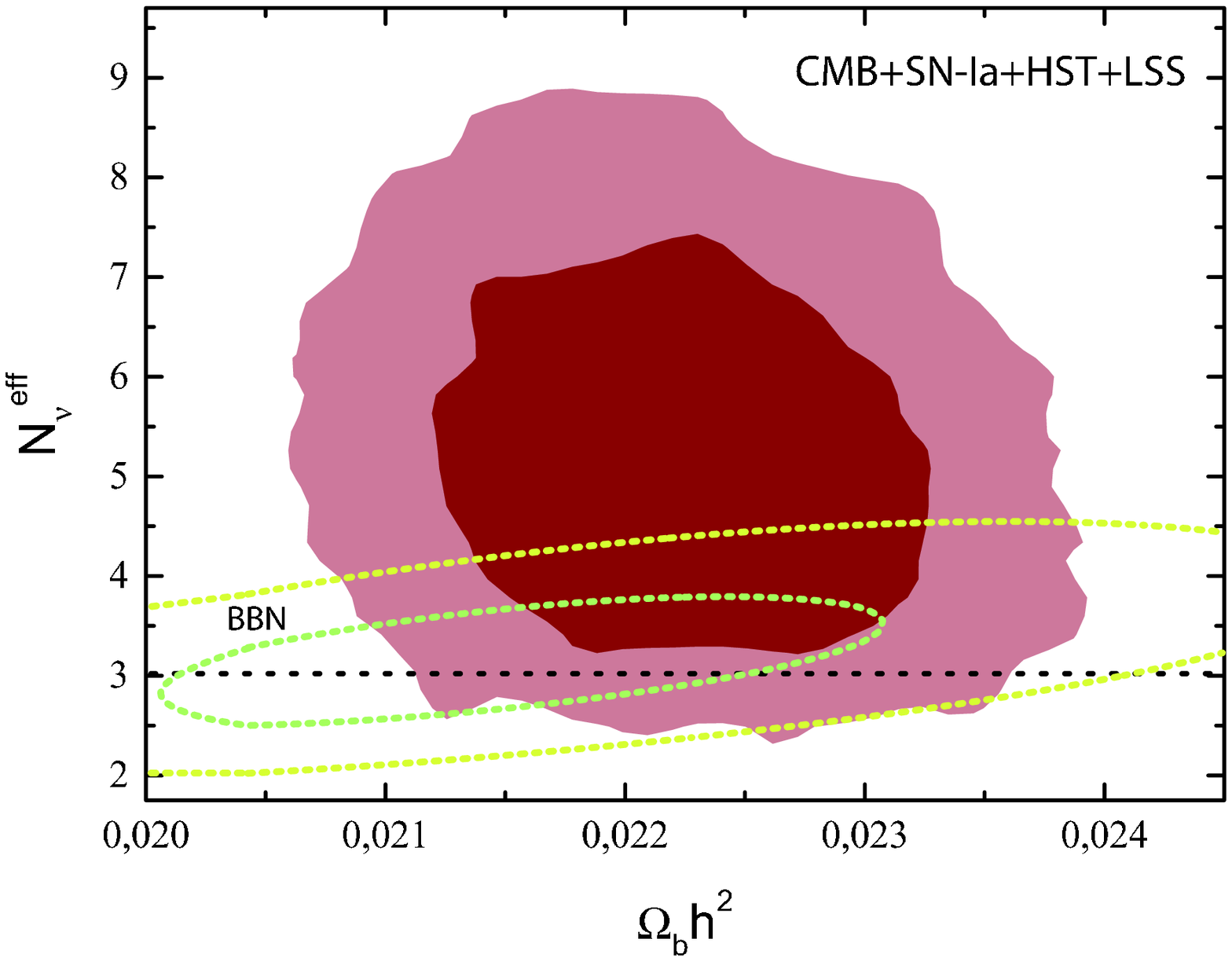} & \includegraphics[width=8.0cm]{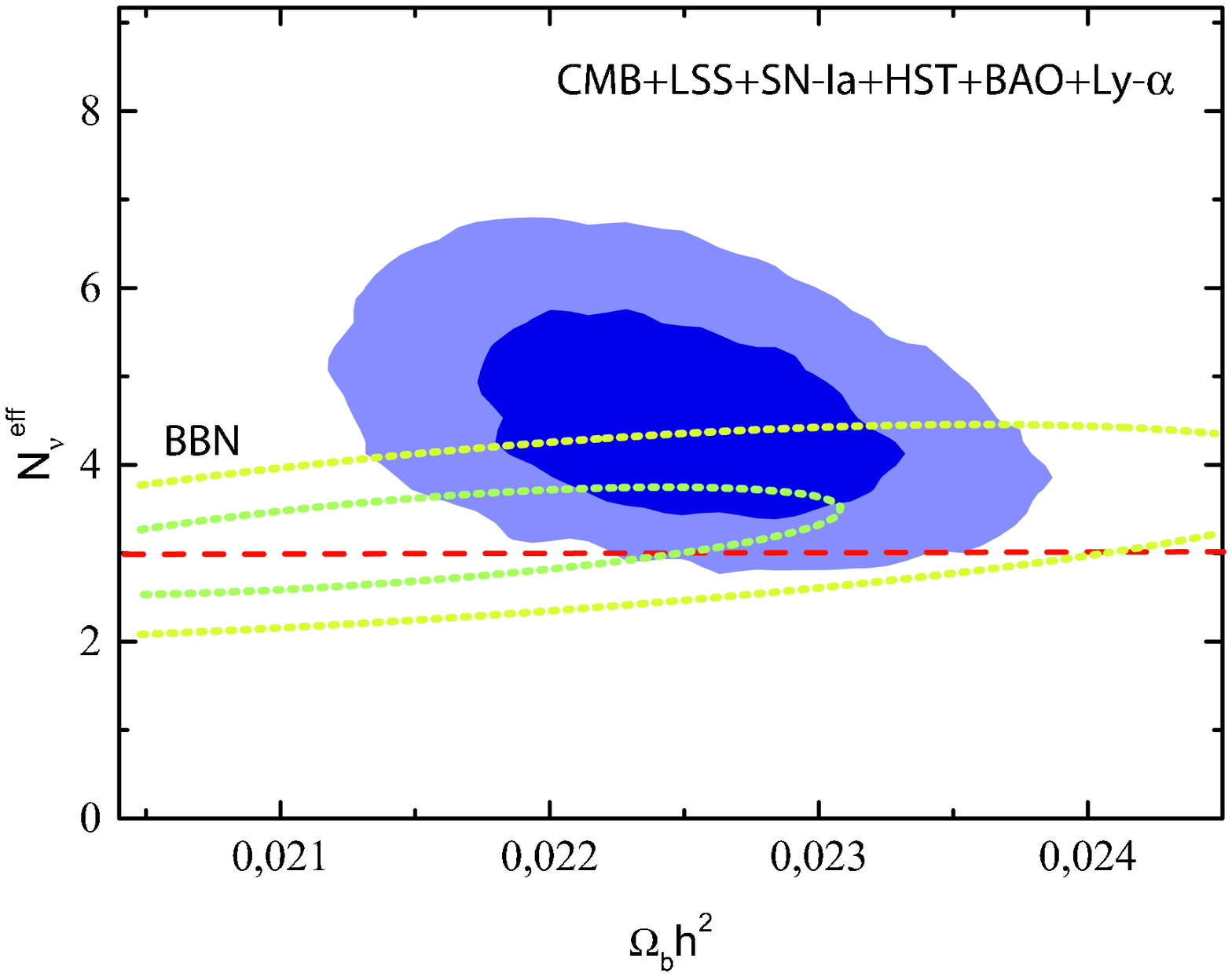}
\end{tabular}
\caption{\label{fig1} Analysis from the first (more conservative)
set of cosmological data (Left Panel) and including Lyman-$\alpha$
and BAO data (Right Panel). We show the marginalized contours at
$68 \%$ and $95 \%$ c.l. on the $\omega_b $-$N_{\nu}^{\rm \it eff}$ plane
along with the analogous contours from BBN using D and $^4$He
(cnservative) experimental results reported in Eqs. (\ref{deut})
and (\ref{4he}) (dotted lines).}
\end{figure}

In Figure \ref{fig1} we show the marginalized constraints at $68
\%$ and $95 \%$ c.l. in the $\omega_b$-$N_{\nu}^{\rm \it eff}$ plane
obtained using the first dataset (left panel) and the results from
the second datasets which includes Lyman-$\alpha$ data and BAO
(right panel). In both cases we also show the analogous contours
from BBN using the experimental determination of $^4$He and D
discussed in the previous Section.
As we can clearly see, an extra-relativistic component
($N_{\nu}^{\rm \it eff}>3$) is favored by both sets of cosmological data.
More specifically, we obtain the following marginalized
constraints: $\omega_b=0.0224 \pm 0.0012$ and
$N_{\nu}^{\rm \it eff}=5.2_{-2.2}^{+2.7}$ at $95 \%$ c.l. from the first
dataset and $\omega_b=0.0224 \pm 0.0011$ and
$N_{\nu}^{\rm \it eff}=4.6_{-1.5}^{+1.6}$ at $95 \%$ c.l. when
Lyman-$\alpha$ and BAO data are included. These data seems to
indicate the presence of an extra-background of relativistic
particles at $2-\sigma$ level, in agreement with the previous
analysis of \cite{Seljak:2006bg} even if with a slightly lower
statistical significance. The small difference from these previous
results is explained by slightly different treatment of the data.
We also note that \cite{Seljak:2006bg} did not use the Hubble
Space Telescope prior by default.

In general, considering a smaller set of data would weaken the
constraints on $N_\nu^{\rm \it eff}$. For example, combining only
the WMAP data with the recent data from the analysis of LRG in the
SDSS \cite{tegmark06} gives $N_\nu^{\rm \it
eff}=3.7_{-2.7}^{+4.4}$. Moreover, using different smaller
datasets may give different bounds on $N_\nu^{\rm \it eff}$. This
is shown in Table \ref{tab1}. Notice that combining WMAP data with
the SDSS power spectrum prefers a large value for $N_\nu^{\rm \it
eff} \sim 8$ though with a large uncertainty, while one gets a
scenario compatible with the standard result $N_\nu^{\rm \it
eff}\sim 3$ if the 2dF data are considered, as already noticed in
previous analyses \cite{Seljak:2006bg}.
\begin{table}
\caption{\label{tab1}Constraints on $N_{\nu}^{\rm \it eff}$ from
different datasets. For each dataset we also show the
corresponding best fit $\Delta \chi^2$ between model with
$N_\nu^{\rm \it eff}=3$ and allowing for variation in $N_\nu^{\rm
\it eff}$ .}
\begin{tabular*}{\textwidth}{@{}l*{15}{@{\extracolsep{0pt plus
12pt}}l}} \br
Dataset& $ N_{\nu}^{\rm \it eff}$ (95 c.l.) & $\Delta \chi^2$\\
\mr
CMB+SN-Ia+HST+LSS (first dataset)& $5.2_{-2.2}^{+2.7}$ & 2.9\\
CMB+LSS+SN-Ia+HST+BAO+Ly-$\alpha$ (second dataset) & $4.6_{-1.5}^{+1.6}$ & 3.8\\
CMB+LRG (SDSS) & $3.7_{-2.7}^{+4.4}$ & 0.6\\
WMAP+2dF & $3.2_{-2.3}^{+3.6}$ & 0.7\\
WMAP+BAO & $4.2_{-2.1}^{+2.7}$ & 0.8\\
WMAP + SDSS & $7.8_{-3.2}^{+2.3}$ & 6.2\\
WMAP+BAO+Ly-$\alpha$+SN-Ia & $5.2_{-1.8}^{+2.1}$ & 4.6 \\
\mr
BBN D \cite{O'Meara:2006mj} + $Y_p=0.249\pm0.009$ \cite{olive} & $3.1_{-1.2}^{+1.4}$ & / \\
BBN D \cite{O'Meara:2006mj} + $Y_p=0.2472 \pm 0.0012$
\cite{Izotov:2007ed} & $3.0 \pm 0.2$ & / \\
BBN D \cite{O'Meara:2006mj} + $Y_p=0.2516 \pm 0.0011$
\cite{Izotov:2007ed} & $3.3 \pm 0.2$ & / \\
\br
\end{tabular*}
\end{table}

On the other hand, BBN is in very good agreement with the standard
result $N_{\nu}^{\rm \it eff} \sim 3$. Using the $D$ and $^4$He mass
fraction of Eqs. (\ref{deut}) and (\ref{4he}), after
marginalization we find $\omega_b=0.0215_{-0.0029}^{+0.0038}$ and
$N_{\nu}^{\rm \it eff}=3.1_{-1.2}^{+1.4}$ at $95 \%$ c.l..

Very recently, a new analysis of $^4$He mass fraction has been
performed using a large and homogeneous sample of extragalactic H
II region \cite{Izotov:2007ed} and taking into account all main
source of systematics via a Monte -- Carlo approach. This results
in a much smaller uncertainty than in \cite{olive}. Indeed, using
two different He I line emissivities, they find $Y_p=0.2472 \pm
0.0012$ and $Y_p=0.2516 \pm 0.0011$, respectively. These values
when combined with the result on Deuterium of Eq. (\ref{deut})
lead to a much more constrained bound on the effective neutrino
number, $N_{\nu}^{\rm \it eff}= 3.0 \pm 0.2$ and $N_{\nu}^{\rm \it
eff}= 3.3 \pm 0.2$ at 95 c.l.. Marginalization over $N_{\nu}^{\rm
\it eff}$ gives a basically unchanged result for the baryon
fraction, $\omega_b=0.0211_{-0.0020}^{+0.0028}$ and
$\omega_b=0.0217_{-0.0021}^{+0.0028}$. All constraints from BBN
are also summarized in Table \ref{tab1}.

Notice that all cosmological data, spanning several decades in
redshift or photon temperature provides results which are in
agreement at the $2-\sigma$ level, with the only exception of the
WMAP+SDSS dataset which singles out a very large value of
$N_{\nu}^{\rm \it eff}$, see Table \ref{tab1}. We conclude that
presently there is not a strong evidence for different values of
$N_{\nu}^{\rm \it eff}$ at the BBN and late epochs and that a
slightly non standard value $N_{\nu}^{\rm \it eff}\sim 4$ seems to
be preferred when combining all cosmological observables (first
two rows of Table \ref{tab1}).

\begin{figure}
\begin{tabular}{cc}
  \includegraphics[width=7.1cm]{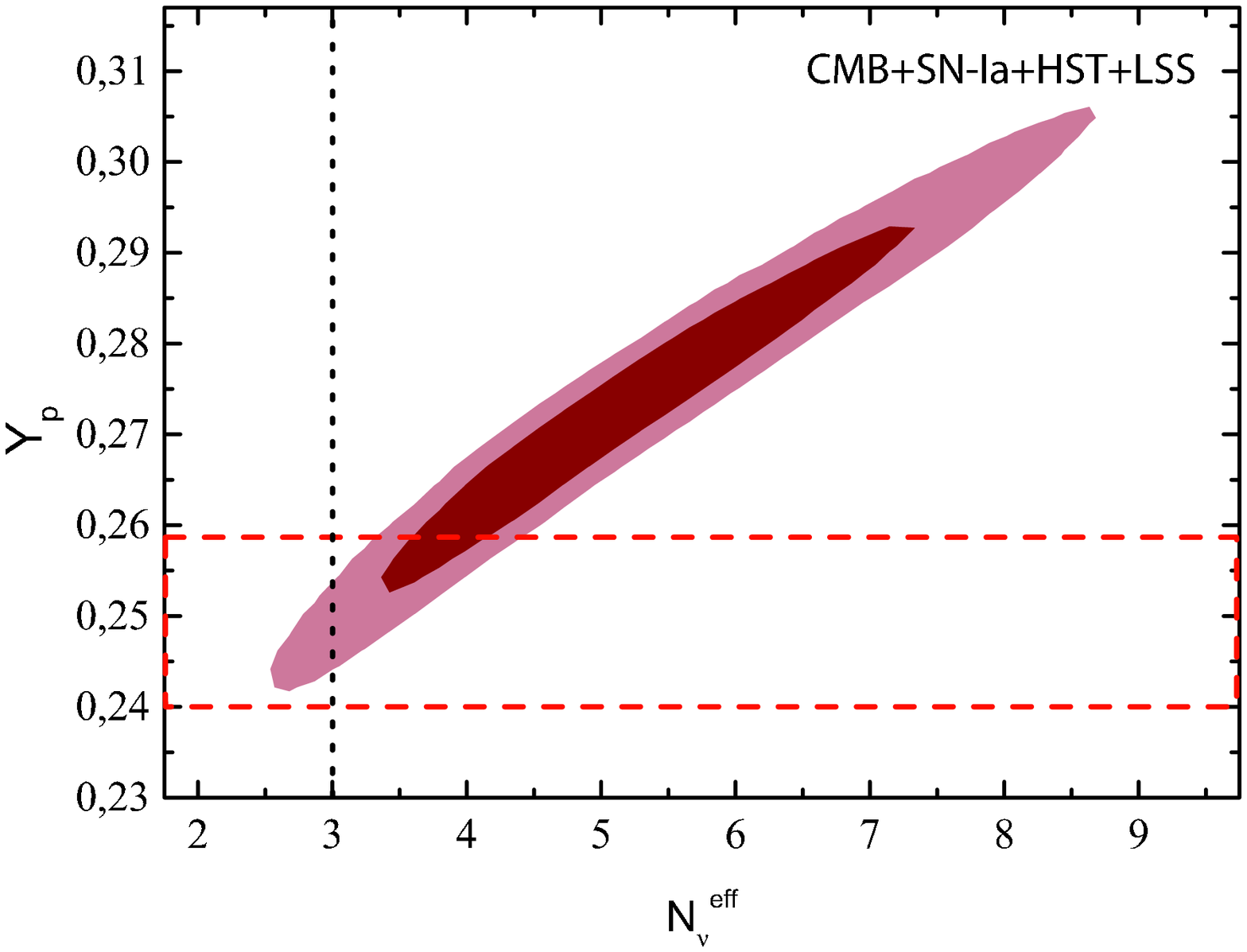} & \includegraphics[width=7.0cm]{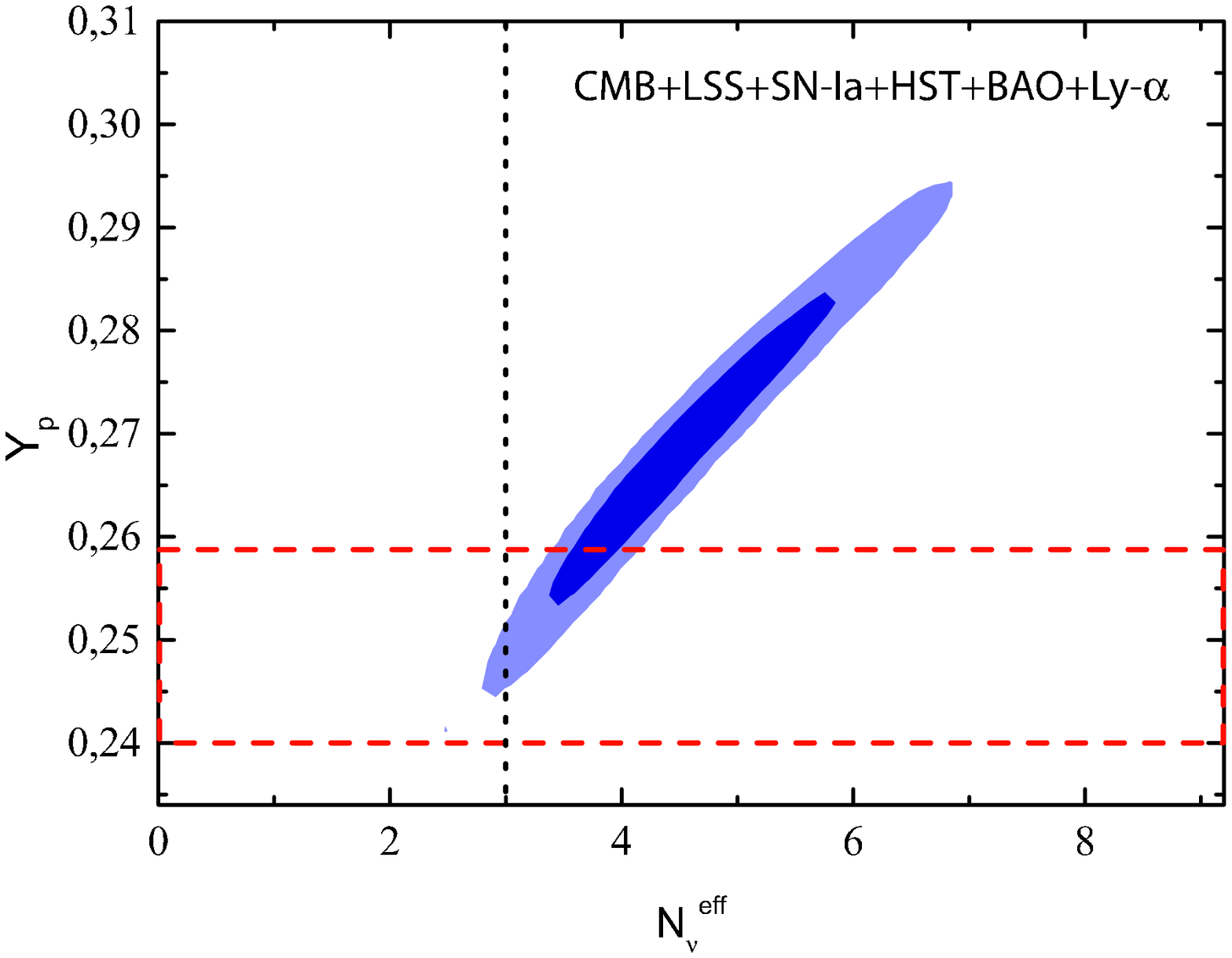} \\
\includegraphics[width=7.1cm]{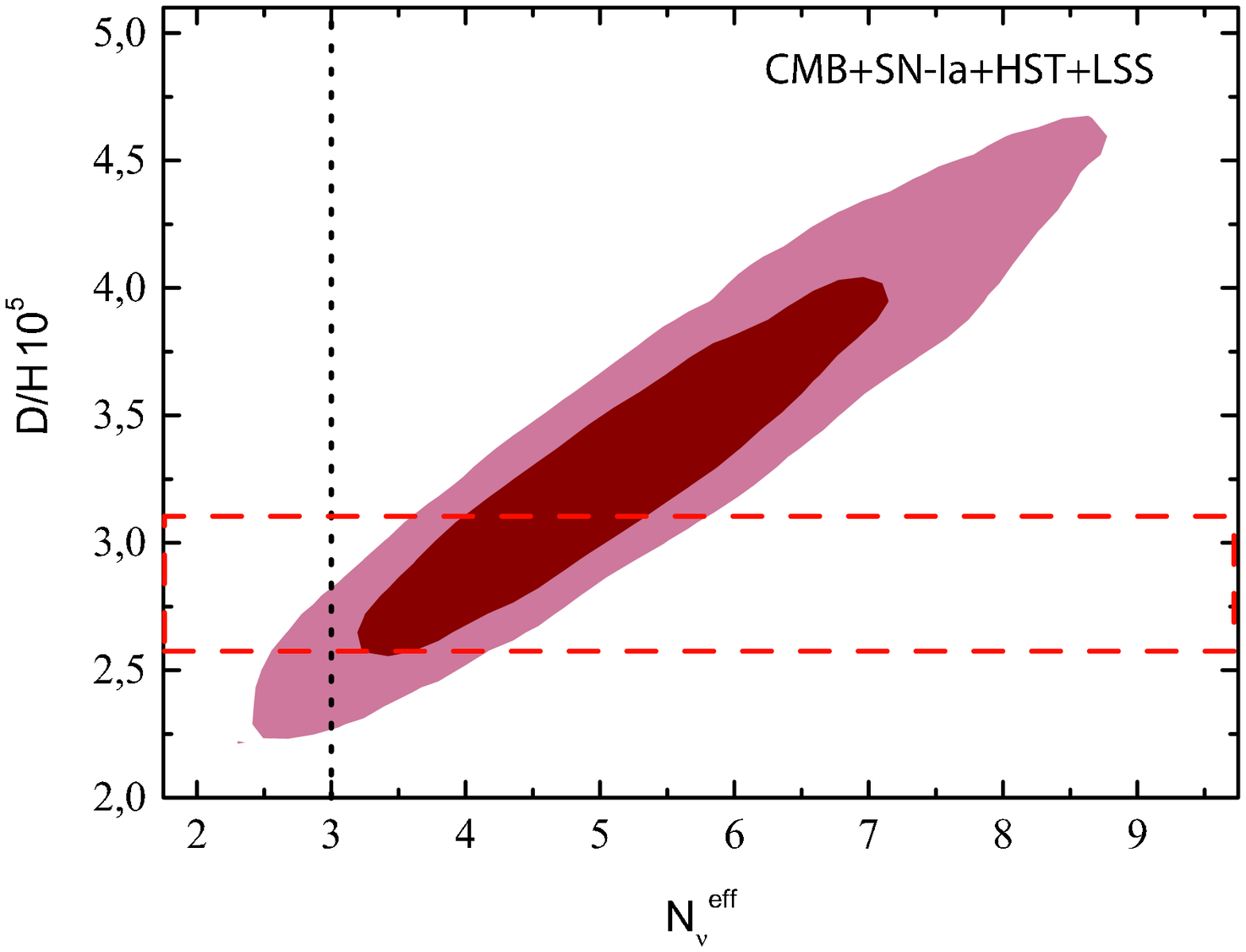} & \includegraphics[width=7.0cm]{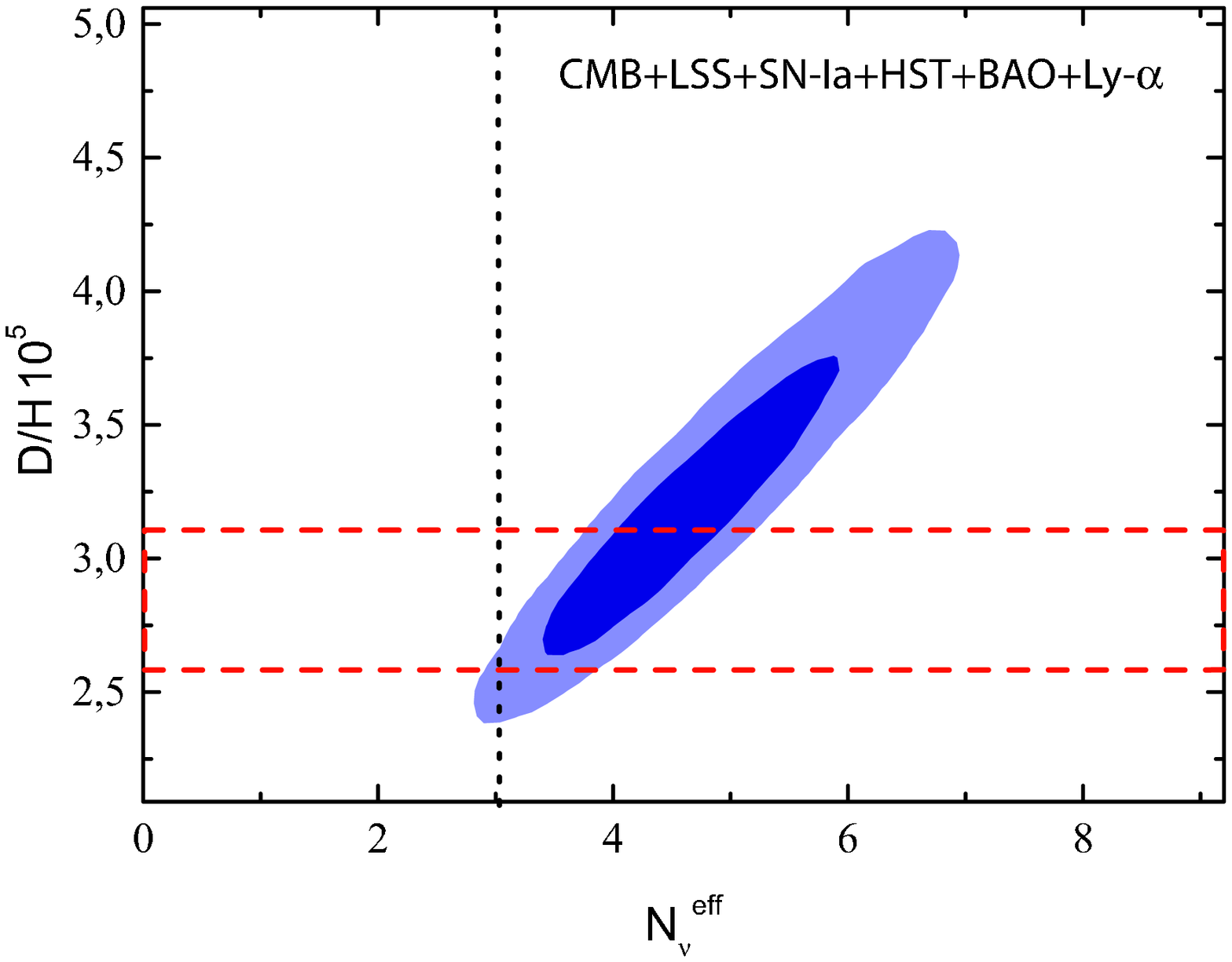}
\end{tabular}
\caption{\label{fig2} The $^4$He mass fraction (Top left Panel)
and D/H ratio (Bottom left Panel) from the first (more
conservative) set of cosmological data versus $N_{\nu}^{\rm \it eff}$ (see
text). We show the marginalized constraints at $68 \%$ and $95 \%$
c.l., while the regions between the two dashed horizontal lines
are the 1-$\sigma$ experimental measurement bands. Plots on the
right show the same results but adding BAO and Lyman-$\alpha$
data}
\end{figure}

It is also interesting to present our results in a slightly
different way, by considering for each nuclide abundance its
dependence on $N_{\nu}^{\rm \it eff}$. Assuming that the value of
$N_{\nu}^{\rm \it eff}$ {\it does not change} between the BBN and the late
CMB and LSS formation epochs we can compute using our numerical
BBN code what are the values of $Y_p$ and D which are preferred at
68 \% and 95 \% c.l. from the results obtained for $\omega_b$ and
$N_{\nu}^{\rm \it eff}$ from the two datasets of the previous Section.
This is shown in Figure \ref{fig2}, along with the corresponding
experimental values. Notice that as well known, both nuclei show a
monotonically increasing behavior as function of the relativistic
energy density. As from Figure \ref{fig1}, we see that the
measured values of $Y_p$ and D/H are in agreement with the
cosmological constraints and a better agreement is obtained for
the non standard value $N_{\nu}^{\rm \it eff}\sim 4$. In particular we can
derive the following marginalized ranges
$Y_p=0.273_{-0.025}^{+0.027}$ and D/H$=(3.4_{-0.9}^{+1.0}) \cdot
10^{-5}$ at $95 \%$ c.l. for the first dataset and
$Y_p=0.266_{-0.020}^{+0.020}$ and D/H$=(3.1_{-0.7}^{+0.8}) \cdot
10^{-5}$ in case we include in the analysis both BAO and
Lyman-$\alpha$ data. As we mentioned, these results are only
meaningful under the assumption that $N_{\nu}^{\rm \it eff}$ does not
change from the BBN till very recent epochs. A different
conclusion would be obtained if we reason in the framework of
models where $N_{\nu}^{\rm \it eff}$ can change in time as those
summarized in Section I. In this case, data shows that nuclei
abundances are best fit by a standard value $N_{\nu}^{\rm \it eff}\sim 3$
at the BBN epoch, while CMB and LSS prefer larger values due to
some extra relativistic degrees of freedom, though with quite a
large error.

%
%

\begin{figure}
\begin{tabular}{cc}
  \includegraphics[width=7.1cm]{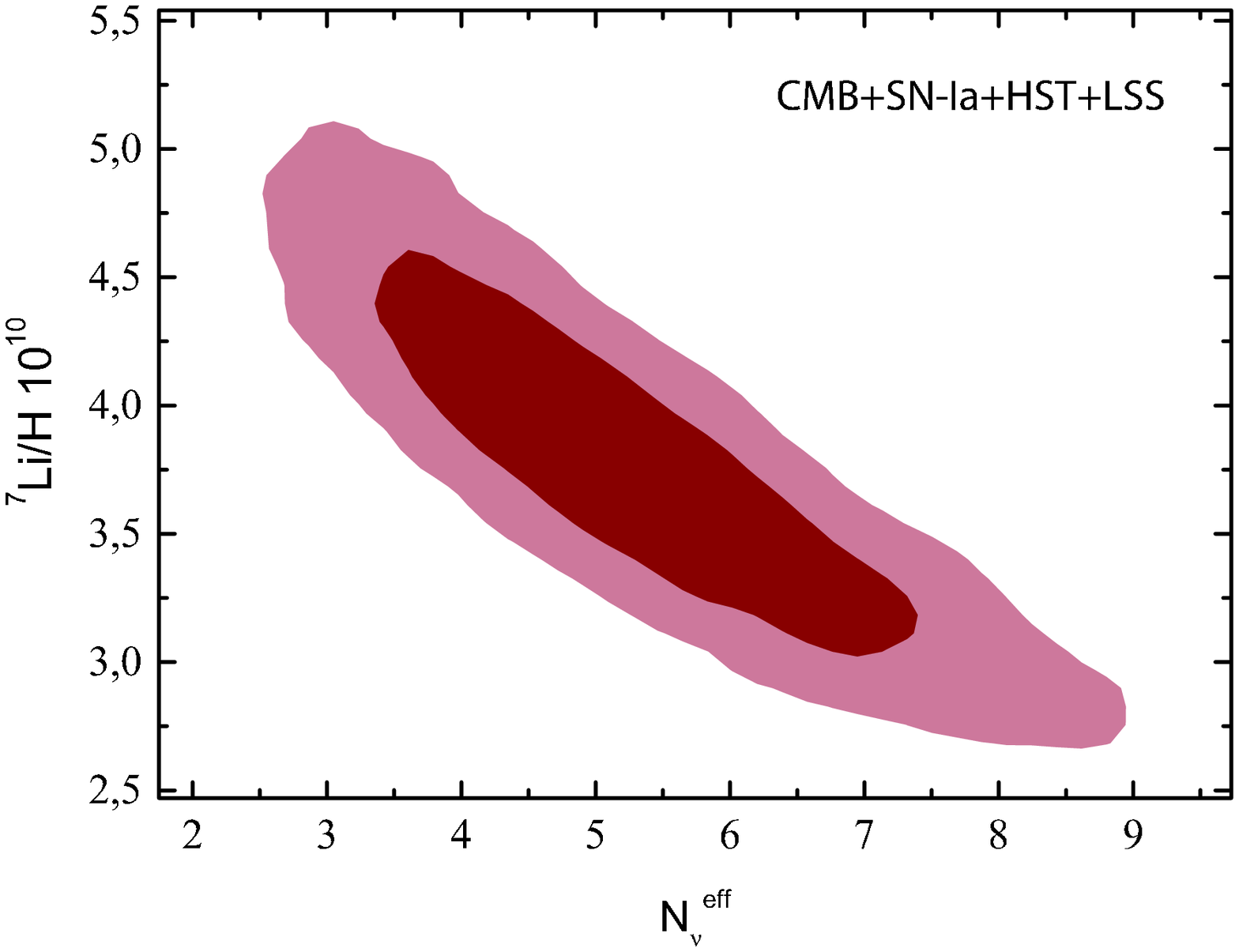} & \includegraphics[width=7.0cm]{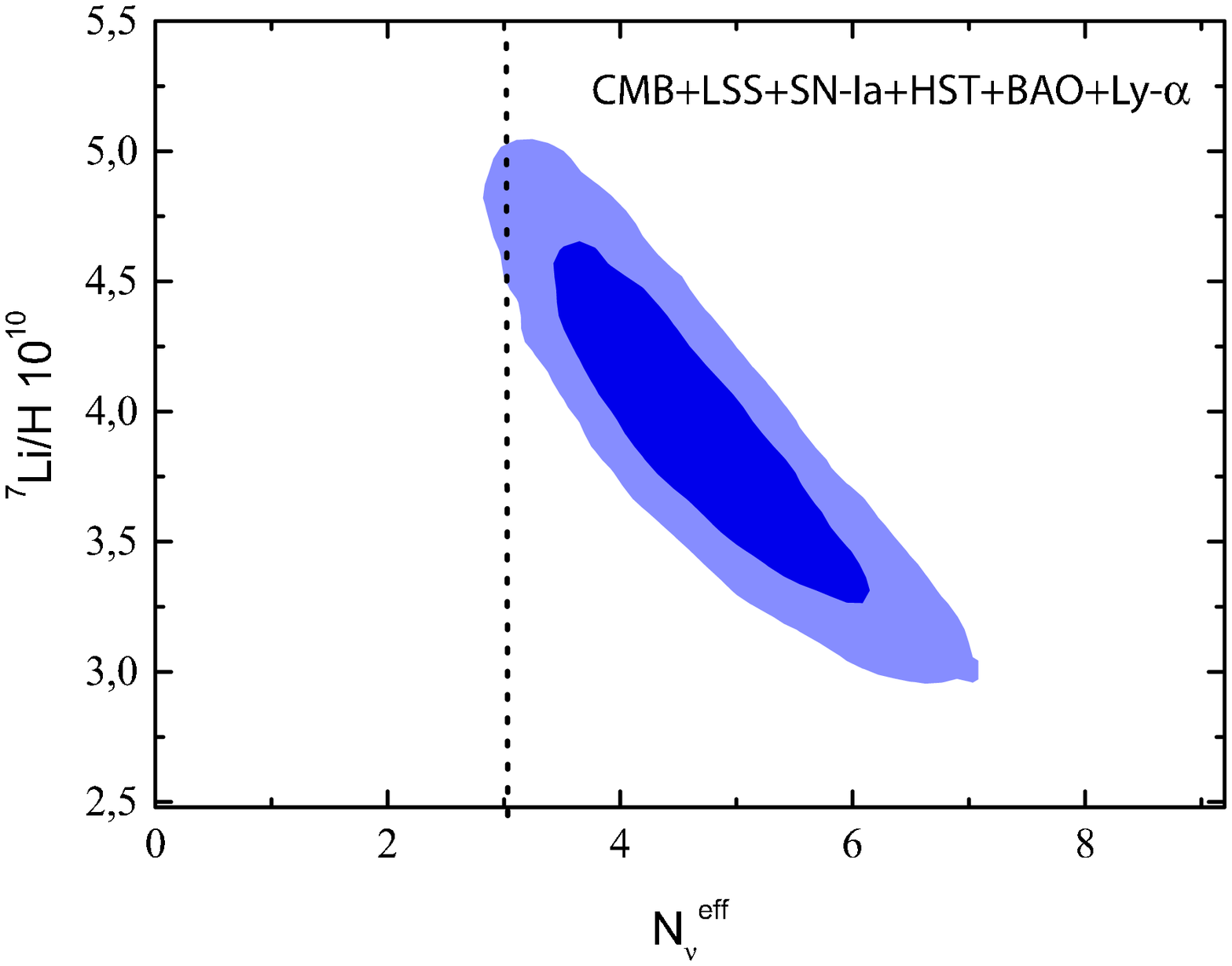} \\
\includegraphics[width=7.1cm]{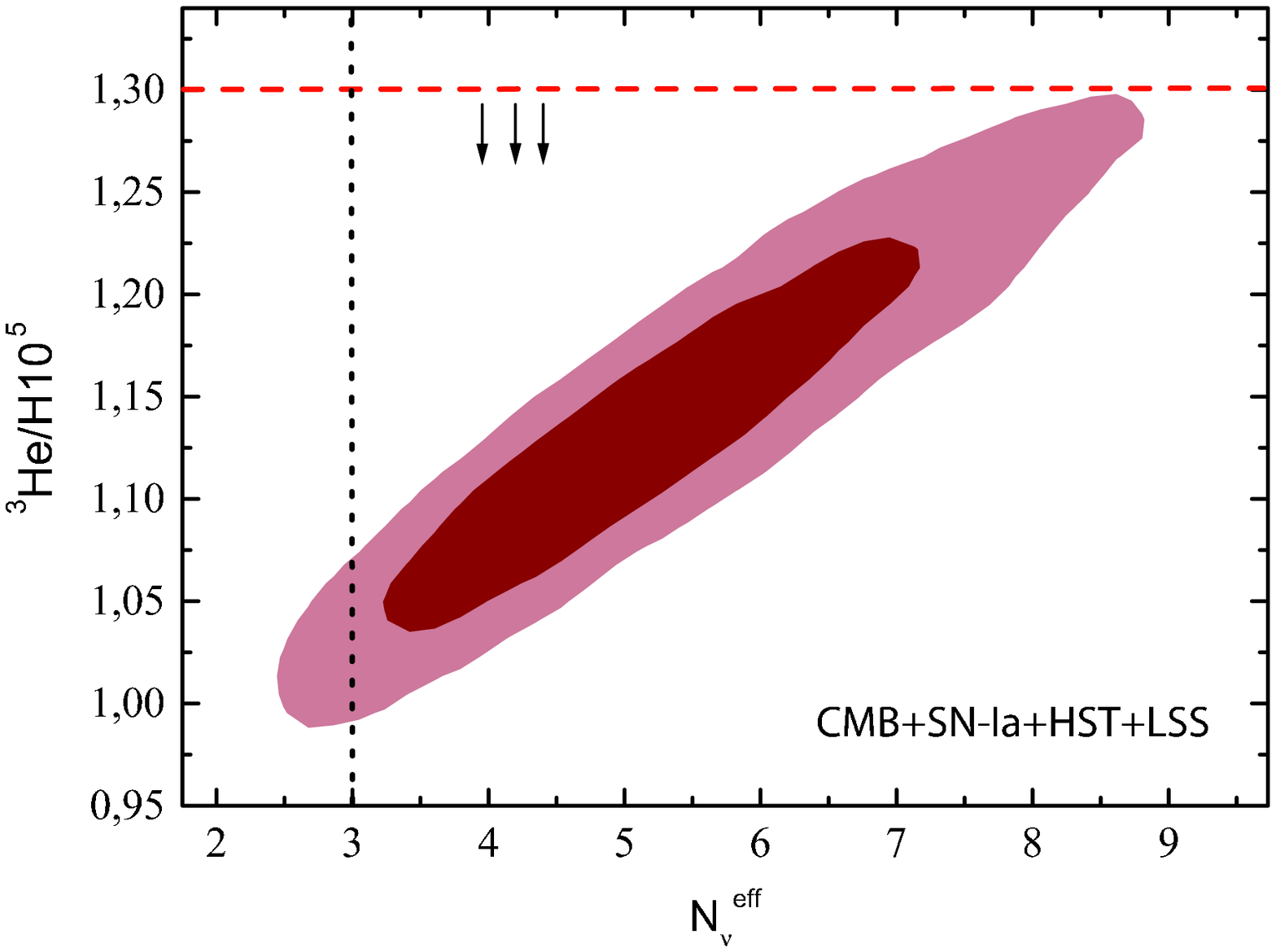} & \includegraphics[width=7.0cm]{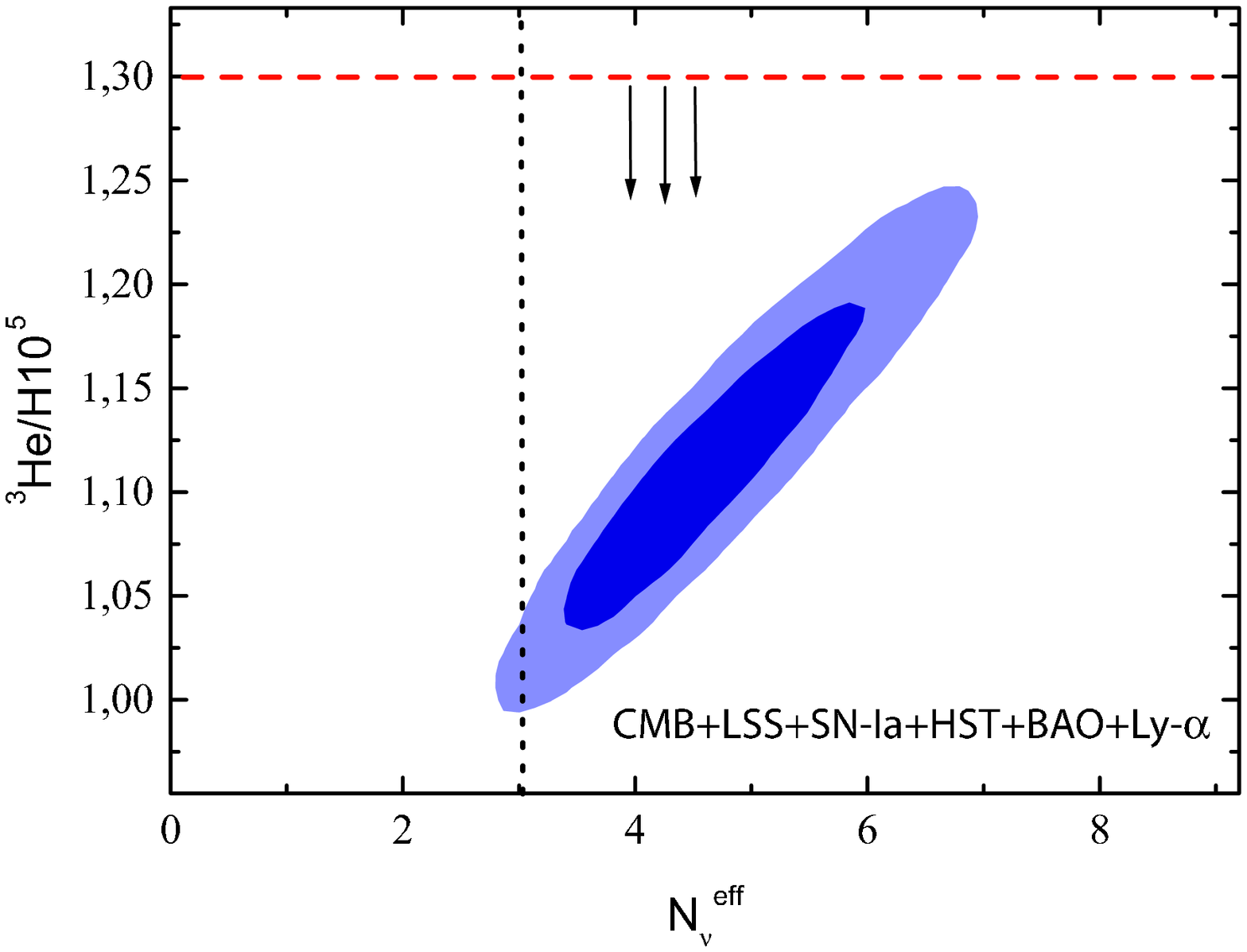}
\end{tabular}
\caption{\label{fig3} The $^7$Li/H (Top Left Panel) and $^3$He/H
(Bottom Left Panel) from the first (more conservative) set of
cosmological data versus $N_{\nu}^{\rm \it eff}$. We show the marginalized
constraints at $68 \%$ and $95 \%$ c.l.. We also show the
1-$\sigma$ experimental upper limit on $^3$He/H of Eq.
(\ref{bound3he}). The experimental results on $^7$Li/H are out of
the chosen range. Plots on the right show the same results but
adding BAO and Lyman-$\alpha$ data}
\end{figure}

Besides $^4$He and D, the most abundant nuclides produced during
BBN are $^3$He and $^7$Li. As for the latter, there are two recent
measurements, namely \be ^7\textrm{Li}/\textrm{H} =(1.23 \pm 0.06)
\cdot 10^{-10} \vv \label{li71} \ee from \cite{ryan} and \be
^7\textrm{Li}/\textrm{H} =(1.26_{-0.24}^{+0.29}) \cdot 10^{-10}
\pp \label{li72} \ee from \cite{bonifacio}. Both these values are
a factor larger than three smaller than what is expected from a
standard BBN scenario with only three active neutrinos
contributing to $N_{\nu}^{\rm \it eff}$, which gives $^7$Li/H=$(4.7 \pm
0.4 ) \cdot 10^{-10}$ for $\omega_b=0.0224$, where the uncertainty
is due to propagation of uncertainties on all nuclear rates
\cite{Serpico:2004gx}. It is quite difficult that the theoretical
estimate may be reconciled with the experimental value, even
allowing for unlikely large values of $N_{\nu}^{\rm \it eff}$ which would
be in striking disagreement with other nuclei abundances. This is
shown in Figure \ref{fig3} (Top Panels) where we report the $^7$Li
abundance versus $N_{\nu}^{\rm \it eff}$ for the two datasets considered
here. Even if $N_{\nu}^{\rm \it eff}$ is of order of 8 - 9 which is
completely excluded by D and $^4$He measurements, the predicted
value for this nuclide is too high, suggesting a strong depletion
mechanism of the Spite plateau value \cite{bonifacio}. This is
also suggested by the fact that a positive dependence on the
metallicity has been found in \cite{ryan}, while there is only a
weak evidence for such a dependence in the star sample studied in
\cite{bonifacio}. In any case, it is quite problematic to compare
BBN predictions with $^7$Li data, and further studies seem
necessary in understanding possible source of systematics in POP
II stellar modelling.
%
%

The results for $^3$He using the two datasets are shown in the
bottom panels of Figure \ref{fig3}. We recall that presently only
an upper bound has been found for the primordial abundance of this
nuclide \cite{bania} \footnote{A weaker bound is obtained by
combining a larger sample of galactic HII regions, $^3$He/H $\leq
(1.9 \pm 0.6) \cdot 10^{-5}$ \cite{bania}.} \be
^3\textrm{He}/\textrm{H} <(1.1 \pm 0.2) \cdot 10^{-5} \vv
\label{bound3he} \ee since the experimental study is highly
complicated by the fact that the primordial $^3$He is strongly
changed by stellar activity \cite{Eggleton:2006uc}. It is
interesting to note however, that this upper bound is very close
to the expected primordial value for a standard BBN scenario.
Using $\omega_b=0.0224$ and three standard neutrinos contributing
to the relativistic energy density we get in fact from the BBN
numerical code used in our analysis $^3$He/H=$(1.02 \pm 0.03)
\cdot 10^{-5}$, where again the uncertainty is due to the errors
in the (experimental) values of the nuclear rates, largely
dominated by the uncertainty in the $^3$He destruction process
$^3$He(d,p)$^4$He \cite{Serpico:2004gx}. By looking at Figure
\ref{fig3} we see that any further improvement in the upper limit
or an experimental estimate for its primordial abundance would
have a big impact in constraining the value of $N_{\nu}^{\rm \it eff}$ at
the BBN epoch. For example, if the upper limit would decrease by
20 \% this may represent a robust evidence in favor of a standard
value $N_{\nu}^{\rm \it eff}\sim 3$ at BBN, so that any indication of
larger values for this parameter by other cosmological observables
would be a strong indication of a late production of relativistic
degrees of freedom {\it after} BBN. From our analysis we also find
that combining all data used so far we can derive the following
conservative lower bound at 95 \% c.l. \be
^3\textrm{He}/\textrm{H} \geq 0.95 \cdot 10^{-5} \vv \ee where we
have also included the effect of the nuclear rate uncertainty in
the theoretical estimate.

\section{Conclusions and Outlooks}

In this paper we have presented an updated analysis of the
constraints on the relativistic energy density in the Universe
combining several cosmological data which {\it measure} the value
of the effective number of neutrinos $N_{\nu}^{\rm \it eff}$ at different
epochs in the evolution of the Universe. Nuclide abundances
produced during BBN can tell us in fact how much energy was in the
form of relativistic species when the photon temperature was in
the range 0.01-1 MeV, while CMB and LSS data provide information
at much later epochs. In the standard scenario after $e^+-e^-$
annihilation only three chiral active neutrinos contribute to
$N_{\nu}^{\rm \it eff}$, and any observed deviation from the value
$N_{\nu}^{\rm \it eff}=3.046$ predicted in the framework of the Standard
Model interactions would be a signal of new physics.

Using WMAP and small scale CMB measurements, LSS and type Ia
Supernova Surveys we find that $N_{\nu}^{\rm \it eff}\sim 3$ is allowed at
2-$\sigma$ though the data prefer somehow larger values. In
particular we obtain $N_\nu^{\rm \it eff}= 5.2^{+2.7}_{-2.2}$ at 95 \%
c.l.. Including Lyman-$\alpha$ and BAO data do not substantially
change the result, $N_\nu^{\rm \it eff}= 4.6^{+1.6}_{-1.5}$ at 95 \% c.l..
On the other hand, data on primordial D and $^4$He are in
excellent agreement with $N_\nu^{\rm \it eff} \sim 3$. This tension
between BBN result and other observations might be interpreted as
a 2-$\sigma$ evidence of the fact that further relativistic
species are produced during the evolution of the Universe after
BBN, but our viewpoint is that this conclusion is still
statistically very weak and further data are necessary.

Combining all data, under the assumption that the number of
relativistic species does not change from BBN down to CMB and LSS
formation epochs, the three neutrino scenario is still quite
consistent, while data show a better agreement for slightly larger
values $N_\nu^{\rm \it eff}= 3-4$. Considering a smaller set of data would
weaken the constraints on $N_\nu^{\rm \it eff}$. For example, we mentioned
that combining only the WMAP data with the recent data from the
analysis of LRG in the SDSS \cite{tegmark06} gives the constrain
$N_\nu^{\rm \it eff}=3.7_{-2.7}^{+4.4}$ with the standard value
$N_\nu^{\rm \it eff}=3.046$ well consistent with the data (see also
\cite{ichikawa}). Moreover, choosing different smaller dataset
also affects the bound on $N_\nu^{\rm \it eff}$, see Table \ref{tab1}.
Finally, we also point out that we assumed in our analysis only
massless neutrinos. The datasets we considered in this paper and
neutrino oscillation experiments are indeed in agreement with a
negligible (for cosmology) neutrino mass (see \cite{Seljak:2006bg}
and \cite{fogli06}). However, considering massive neutrinos would
shift our constraints to even higher values for $N_\nu^{\rm \it eff}$ (see
e.g. \cite{hannestud}).

Apart from further data from CMB anisotropy and polarization which
can be foreseen and available in the near future, as for example
from the \textrm{PLANCK} experiment\footnote{Forecast analysis
estimate a sensitivity of \textrm{PLANCK} on $N_\nu^{\rm \it eff}$ at the
level of 0.2-0.4 , see e.g. \cite{Perotto:2006rj}.}, as well as
further studies on both Lyman-$\alpha$ and BAO data, we stress
that new results on primordial nuclei abundance are needed. The
large (systematic) error on the $^4$He mass fraction is severely
limiting the accuracy of any estimate of $N_\nu^{\rm \it eff}$ at the MeV
scale, while more accurate data on Deuterium, though always
welcome, are not expected to change the overall picture in a
significant manner, unless the accuracy in predicting the baryon
density would become comparable with the one obtained with CMB
measurements. This requires in particular, a better understanding
of the apparent spread of individual D/H measurements in different
QSO systems as due to a real non homogeneous distribution of
primordial Deuterium in the sky or an underestimated statistical
uncertainty in the measurements.

Interestingly, an improved upper bound or measurement of
primordial $^3$He would have a large impact in further
constraining $N_\nu^{\rm \it eff}$. This is unfortunately, a very
difficult task, because of the large contamination of primordial
$^3$He by stellar activity. However, in the framework of
homogeneous BBN there is a robust lower limit on this nuclide
abundance, $^3$He/H $\geq 0.95 \cdot 10^{-5}$, which is very close
to the present experimental upper bound, so we can optimistically
expect some further development in the near future.

\ack

G. Mangano is pleased to thank Galileo Galilei,
Florence, Italy, for the hospitality and the INFN for partial
support during the completion of this work. O. Mena is supported
by the European Programme MRTN-CT-2004-503369. G. Mangano and G.
Miele are supported, in part, by PRIN04 of the Italian MIUR under
grant {\it Fisica Astroparticellare}. P.D. Serpico is gratefully
acknowledged for useful discussions.

\end{document}